\documentclass[preprint,journal]{vgtc}       

\ifpdf
  \pdfoutput=1\relax                   
  \usepackage{graphicx}                
  \DeclareGraphicsExtensions{.pdf,.png,.jpg,.jpeg} 
\else
  \usepackage{graphicx}                
  \DeclareGraphicsExtensions{.eps}     
\fi%

\graphicspath{{figures/}{pictures/}{images/}{./}} 

\usepackage{booktabs}
\usepackage{textcomp} 
\usepackage{microtype}                 
\PassOptionsToPackage{warn}{textcomp}  
\usepackage{textcomp}                  
\usepackage{mathptmx}                  
\usepackage{times}                     
\usepackage{cite}                      
\usepackage{tabu}                      
\usepackage{booktabs}                  

\usepackage{braket}
\usepackage{amssymb}
\usepackage{lipsum}
\usepackage{enumitem}
\usepackage{setspace}

\usepackage[normalem]{ulem}
\usepackage[hyphens]{url}
\useunder{\uline}{\ul}{}

\usepackage{float}
\usepackage{stfloats}

\usepackage{fixltx2e}

\usepackage{caption} 
\usepackage{xcolor}

\usepackage[linewidth=0pt,linecolor=blue]{mdframed}
\newcommand{\modify}[1]{\textcolor{black}{#1}}
\newcommand{\modifyRed}[1]{\textcolor{black}{#1}}

\usepackage{tikz}
\usepackage{xcolor}
\definecolor{lightGrey}{rgb}{0.94, 0.97, 1.0}

\definecolor{component}{RGB}{89, 89, 89}
\newcommand*\component[1]{\tikz[baseline=(char.base)]{
            \node[shape=circle,draw=component,text=component, thick, inner sep= 1pt] (char) {\textsf{#1}}}}
            
\definecolor{subcomponent}{RGB}{137, 137, 137}
\newcommand*\subcomponent[1]{\tikz[baseline=(char.base)]{
            \node[shape=circle,fill=subcomponent,text=white, inner sep= 0.5pt] (char) {\textsf{\small #1}}}}

\definecolor{subsubcomponent}{RGB}{242, 242, 242}
\definecolor{color2}{RGB}{102,102,102}

\definecolor{redSquare}{RGB}{255, 92, 15}
\newcommand*\redSquare[1]{\tikz[baseline=(char.base)]{
            \node (rect) [fill=redSquare,thick, inner sep= 3pt] {}}}
            
\definecolor{blueSquare}{RGB}{8, 174, 255}
\newcommand*\blueSquare[1]{\tikz[baseline=(char.base)]{
            \node (rect) [fill=blueSquare,thick, inner sep= 3pt] {}}}

\newcommand*{\rom}[1]{\expandafter\@slowromancap\romannumeral #1@}

\newcommand{\toolName}{\textit{VACSEN}}

\onlineid{1351}

\vgtccategory{Research}

\vgtcinsertpkg

\title{\toolName{}: A \underline{V}isualization \underline{A}pproa\underline{c}h for Noi\underline{s}e Awaren\underline{e}ss in Qua\underline{n}tum Computing}

\author{Shaolun Ruan, Yong Wang, Weiwen Jiang, Ying Mao, and Qiang Guan}
\authorfooter{
\item
 S. Ruan and Y. Wang are with Singapore Management University. Yong
 Wang is the corresponding author. E-mail: slruan.2021@phdcs.smu.edu.sg and 
 yongwang@smu.edu.sg.
\item
 W. Jiang is with George Mason University. E-mail: wjiang8@gmu.edu.
\item
 Y. Mao is with Fordham University. E-mail: ymao41@fordham.edu.
\item
 Q. Guan is with Kent State University. E-mail: qguan@kent.edu.
}

\teaser{
  \centering
  \includegraphics[width=0.95\linewidth]{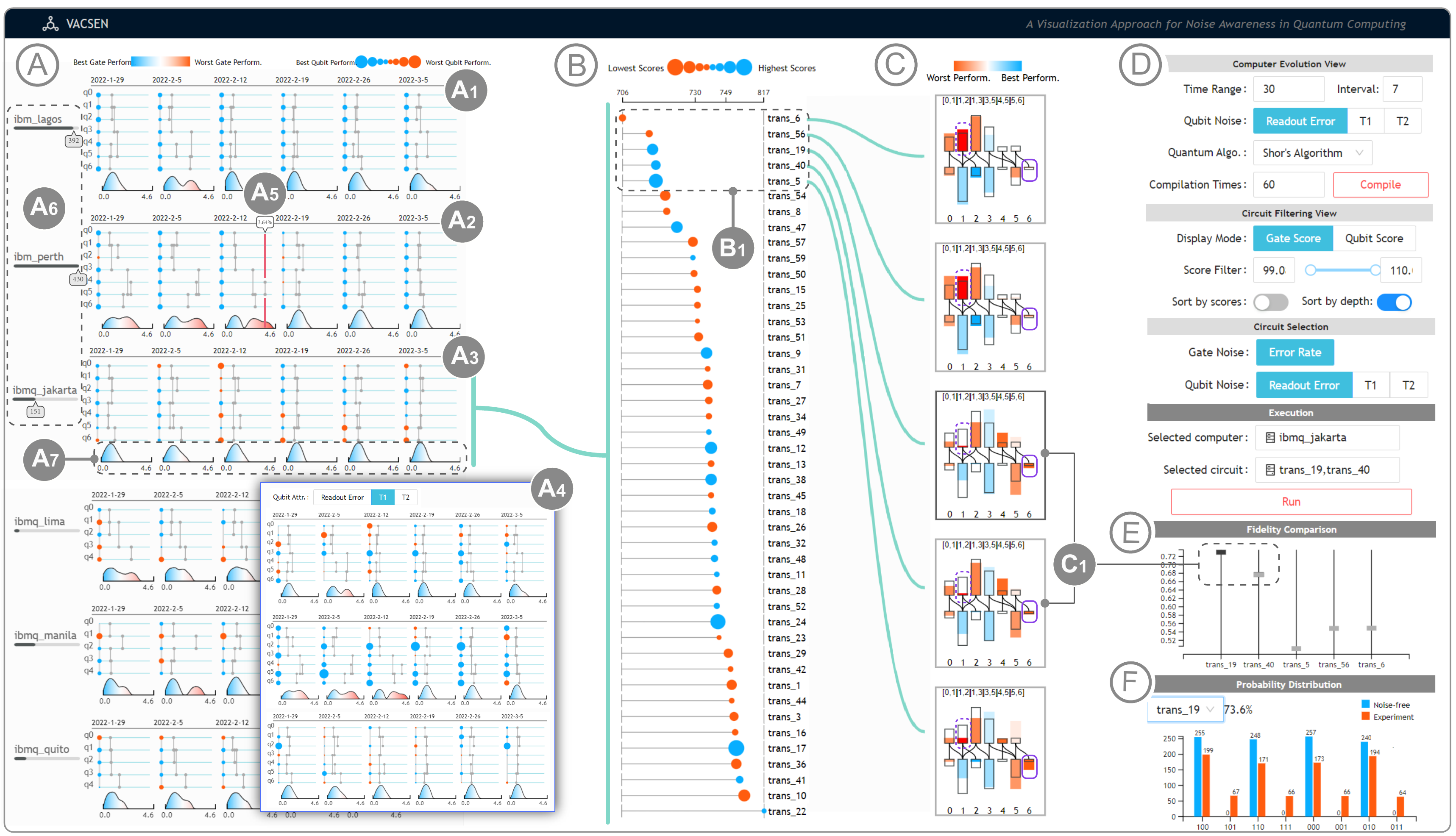}
  \caption{The interface of \toolName\ makes users aware of the quantum noise via three linked views (A-C). Computer Evolution View (A) allows the assessment for all quantum computers based on a temporal analysis for multiple performance metrics. Circuit Filtering View (B) supports the filtering for the potential optimal compiled circuits. Circuit Comparison View (C) supports the in-depth comparison regarding the performance of qubits or quantum gates and corresponding usages. \modify{
  The control panel (D) allows users to interactively configure the settings of \toolName.
  Fidelity Comparison View (E) shows the fidelity distribution of each compiled circuit.
  Probability Distribution View (F) visualizes the results of state distribution of a quantum circuit execution.
  }}
  \label{fig:teaser}
}

\abstract{Quantum computing has attracted considerable public attention due to its exponential speedup over classical computing. 
Despite its advantages, today's quantum computers intrinsically suffer from noise and are error-prone. 
To guarantee the high fidelity of the execution result of a quantum algorithm, it is crucial to inform users of the noises of the used quantum computer and the compiled physical circuits. 
However, an intuitive and systematic way to make users aware of the quantum computing noise is still missing.
In this paper, we fill the gap by proposing a novel visualization approach to achieve noise-aware quantum computing. 
It provides a holistic picture of the noise of quantum computing through multiple interactively coordinated views:
a Computer Evolution View with a circuit-like design
overviews the temporal evolution of the noises of different quantum computers, a
Circuit Filtering View facilitates quick filtering of multiple compiled physical circuits for the same quantum algorithm, and a Circuit Comparison View with a coupled bar chart enables detailed comparison of the filtered compiled circuits. 
We extensively evaluate the performance of \toolName{} through two case studies on quantum algorithms of different scales and an in-depth interviews with 12 quantum computing users. 
The results demonstrate the effectiveness and usability of \toolName{} in achieving noise-aware quantum computing.
} 


\CCScatlist{ 
 \CCScat{K.6.1}{Management of Computing and Information Systems}%
{Project and People Management}{Life Cycle};
 \CCScat{K.7.m}{The Computing Profession}{Miscellaneous}{Ethics}
}

\begin{document}




\firstsection{Introduction}
\label{sec:intro}
%
\maketitle

%

Along with the evolution of actual quantum computers from industry, like IBM, IonQ, Rigetti, and D-Wave,
quantum computing research has been rapidly growing in the past few years.
An increasing number of
quantum algorithms have been developed and shown a significant speedup over their best-known classical counterparts \cite{shor1999polynomial,kwiat2000grover,arute2019quantum, moller2017impact}.
On top of the breakthroughs in quantum algorithms, quantum computing has become a promising method in many important applications such as
chemistry simulation~\cite{peruzzo2014variational}
and machine learning~\cite{acampora2019quantum, wiebe2014quantum, biamonte2017quantum,jiang2021co}.

Despite the promising
impacts on speeding up algorithms, the high noise level in quantum devices is one of the key challenges to achieve the real quantum advantage in today's quantum computers, which are well known as \textit{Noisy
Intermediate-Scale Quantum (NISQ)} computers.
The noise
originates from a quantum computer's fundamental components~\cite{ preskill2018quantum}, including (1) \textit{qubits}, which are the basic units to store the quantum states; and (2) \textit{quantum gates}, which are the basic operators to manipulate qubits' states.
For example, due to the volatility and limited coherence time, flip errors can occur at qubits~\cite{chuang1995quantum}.

The quantum noise issue becomes more  severe due to the variation of the topology (\textit{i.e.,} connection) on physical qubits.
Specifically, quantum algorithms are synthesized as quantum circuits (also called \textit{logical circuits}~\cite{sheikhfaal2015designing}).
The quantum circuits will finally be deployed to the physical qubits for execution.
It will go through the process of mapping logical circuits to physical qubits to obtain the physical quantum circuits (also called \textit{compiled circuits}~\cite{venturelli2018compiling,tan2020optimality}).
The compiled circuit, however, is specified to the actual quantum computer since the topology of qubits is different.
Thus, even for the same quantum algorithm, the effects of quantum noise on the algorithm can be significantly different.



%

In NISQ era, it is crucial to improve the fidelity of quantum computing, enabling the correctness of quantum algorithms (or at least approximate to the desired results).
To this end, designers of quantum algorithms need to have a clear understanding of the 
hidden noise of 
different quantum computers
as well as the noise inside different compiled circuits on the same quantum computer for a specific quantum algorithm~\cite{shaydulin2019network}.
However, this is not a trivial task.
On the one hand, the access to most of today's quantum computers has been provided by IT companies, like IBM, Amazon, and Microsoft, as a cloud service.
To the best of our knowledge, no uniform tool revealing the noise exists to assist the designers.
On the other hand, 
due to the rapid increase of quantum computing users~\cite{webpageLink}, we face the increasing queuing time for accessing cloud quantum computers without noise awareness.
Thus, the evaluation and mitigation of noise before execution has become even more urgent to avoid the waste of queuing time for unsatisfied results.
Currently, the common practice to obtain less-noisy execution results is still a trial-and-error process.


%
%
%

To address the above problems, it is essential to have a tool to effectively inform users of the noise in quantum computing for a better selection of quantum computers and compiled circuits.
\modify{
Given that visualization has shown great power in various applications~\cite{munzner2014visualization}, we aim to develop a novel visualization approach to enhance users' noise awareness in quantum computing and make quantum computing more transparent and reliable.}
However, there exist many challenges
that mainly come from two perspectives:
\textit{complex and dynamically-evolving quality of quantum computers} and \textit{significant variations of the compiled circuits}.
First, the performance
of quantum computers 
relies on multiple factors of qubits and quantum gates~\cite{saki2021survey, reagor2018demonstration, schlosshauer2019quantum, joos1985emergence}, such as
quantum decoherence, gate error, and qubit readout error.
These factors are dynamically changing over time~\cite{bjfa}.
\modify{It is challenging to visualize these complex factors as well as the qubit topological connections along a timeline.}
Second, 
a quantum algorithm can be compiled to various compiled circuits with significantly-different noise on the same quantum computer.
For a large-scale quantum algorithm, the compiled circuits can be several hundreds~\cite{salm2021automating, bassman2020domain}.
The noise of compiled physical circuit needs to be evaluated from different perspectives, \textit{e.g.}, the circuit depths and the noise of involved qubits or quantum gates~\cite{ash2019qure}.
\modify{But it is difficult to visually summarize a large number of the compiled circuits regarding the various noises, and enable users to select the most appropriate one shortly.}

%
%
%


To fill the research gap, we propose \toolName{}\footnote{\toolName{}
is pronounced as ``vaccine''. We envision \toolName{} can provide insights for designers to vaccinate their quantum algorithms against noise.
}, a \underline{\textbf{V}}isualization \underline{\textbf{A}}pproa\underline{\textbf{C}}h for noi\underline{\textbf{S}}e awaren\underline{\textbf{E}}ss in qua\underline{\textbf{N}}tum computing. \toolName\ can inform
quantum computing users of the noise in quantum computers and compiled physical circuits, leading to a better execution result with higher fidelity.
We follow a user-centered design process~\cite{munzner2009nested} by working closely with five domain experts in quantum computing for over five months.
A pilot study is conducted to derive the design requirements.
These design requirements guide our subsequent visual designs for \toolName{}.
\toolName{} mainly consists of three novel visualization views: Computer Evolution View, Circuit Filtering View, and Circuit Comparison View. 
Specifically, Computer Evolution View (Fig. \ref{fig:teaser}\component{A}) facilitates the temporal noise assessment of quantum computers by a novel circuit-like design that reveals the qubit connectivity.
Circuit Filtering View (Fig. \ref{fig:teaser}\component{B}) supports the filtering of the compiled circuits, allowing users to pick the compiled circuits of interest. 
Circuit Comparison View (Fig. \ref{fig:teaser}\component{C}) further enables a more detailed comparison of selected compiled circuits with a novel coupled bar chart design, facilitating the selection of an optimal compiled circuit for the final execution. 
To the best of our knowledge, \toolName{} is the first visualization approach for 
 real-time noise awareness
in quantum computing.

To evaluate the usefulness and effectiveness of \toolName{}, we present two case studies on both small-scale and large-scale quantum algorithms and conduct in-depth interviews with 12 target quantum computing users.
The results show that \toolName{} can provide users with an intuitive way
to be aware of quantum noise.
The major contributions of this paper can be summarized as follows:

\begin{itemize}
    \item We formulate the design requirements for interactive visual analysis of the noise in quantum computers and compiled circuits, together with quantum computing experts. 
    
    \item We present an interactive visual analytics approach, \toolName{}, to help quantum computing users assess
    the noise in quantum computing. Two novel visual designs are proposed: a circuit-like design facilitates the temporal analysis of quantum computer noise, and a coupled bar chart design enables the in-depth comparison of compiled circuits.
    

    \item We conduct two case studies and in-depth user interviews with expert users to demonstrate the effectiveness and usability of \toolName{}.
\end{itemize}

\section{Related Work}

Our work is relevant to prior research on
visualization for quantum computing and the reliability improvement in quantum computing.

\subsection{Visualization for Quantum Computing}

Researchers have attempted to explain quantum computing via visualization. Miller \textit{et al.}~\cite{miller2021graphstatevis} proposed a node-link approach to present quantum circuits.  Lin \textit{et al.}~\cite{lin2018quflow} introduced a method to reveal the parameters' dynamic change by the sequence of quantum gates. Tao \textit{et al.}~\cite{tao2017shorvis} introduced an interactive platform for users to better understand Shor's algorithm. Also, there are online platforms that support an interactive implementation of circuits. For example, \textit{Quirk}~\cite{quirk} provides a graphical tool to make users aware of quantum circuit's behaviors and state changes. Some commercial vendors also provide cloud platforms that allow users to build quantum circuits interactively, \textit{e.g.}, IBM Quantum\footnote{IBM Quantum: \url{https://www.ibm.com/quantum-computing/}} and Amazon Rigetti\footnote{Rigetti: \url{https://www.riggeti.com/}}.

All of the above studies focused on visualizing qubit states for a better understanding of the quantum workflow. 
\modify{The existing quantum computing platforms (\textit{\textit{e.g.}}, IBM Quantum) often provide users with some quality indicators of each qubit.
However, these platforms can not make users aware of the noise in the compiled circuit, making the circuit execution less reliable.}
Our system aims to
enable multi-level noise awareness
in quantum computing through a novel visualization approach and then help enhance the fidelity of the execution of a quantum algorithm with noise awareness.

\subsection{\modify{Reliability Improvement in Quantum Computing}}

\modify{Reliability improvement studies different methods to enhance the probability of correct results of quantum circuit execution, and has become a prominent research topic in quantum computing~\cite{sodhi2021quantum, liu2021qucloud, thornton2020introduction}.}
There are three major ways to enhance the reliability in quantum computing~\cite{ash2019qure},
\textit{i.e.,} building better qubits,
quantum error correction,
and qubit re-mapping.
To build better qubits, the decoherence time of qubits has been improved significantly in the last two decades~\cite{hou2010suppressing, amin2009decoherence}. However, qubits and gate operations can be made more robust and accurate from a fabrication and control perspective, and there are still some bottlenecks. 
Quantum error correction~\cite{lidar2013quantum}
requires significant qubit overhead (\textit{e.g.}, 18X extra qubits in the surface code~\cite{horsman2012surface}). 
%
%
\modify{Furthermore, quantum computing researchers have studied methods of optimizing the mapping from logical qubits to more reliable physical qubits by considering the noise of physical qubits~\cite{deng2020codar, niemann2020design}}.
Ash-Saki \textit{et al.}~\cite{ash2019qure} provided an approach where high-quality qubits are prioritized for mapping. 
Bhattacharjee \textit{et al.}~\cite{bhattacharjee2019muqut} proposed a mapping approach considering the nearest neighbor compliance. 
Alam \textit{et al.}~\cite{alam2020circuit} studied a qubit mapping algorithm to optimize circuits by exploiting gate re-ordering.



However, existing quantum computers still suffer from noises and the above approaches cannot achieve error-free quantum computing.
Thus, it is still crucial to provide users with an intuitive way to understand and compare the noise of both quantum computers and compiled circuits, which is the focus of this paper.


\section{Background}
\label{sec:background}

This section introduces the overall background of quantum computing which is relevant to our study, including quantum computer architecture, quantum computing workflow, fidelity and noise in quantum computing.

\subsection{Quantum Computer Architecture}

\textbf{Qubit}, or quantum bit,
is the basic unit of quantum information in quantum computing.
For classical computing, a bit always has two deterministic states, \textit{i.e.,} ``0'' or ``1''.
For quantum computing, there are two orthonormal basis states for a qubit, \textit{i.e.,} 
$\ket{0}$ and $\ket{1}$.
The actual quantum state of a qubit can be represented by a linear \textit{superposition} of the two basis states~\cite{brassard1998quantum}:


\begin{equation}
\label{equation:1}
\ket{\Psi} = \alpha\ket{0} + \beta\ket{1},
\end{equation}

where $\alpha, \beta$ are complex numbers subject to $|\alpha|^2 + |\beta|^2 = 1$~\cite{brassard1998quantum}. 
Also, multiple qubits can be entangled to support the qubit interactions, which is called \textit{quantum entanglement}~\cite{horodecki2009quantum}. An entangled state of the two qubits can be made via a gate on the control qubit, followed by the so-called CNOT gate~\cite{alsina2016experimental}.

\textbf{Quantum Gate}, or quantum logical gate,
is a basic quantum circuit operating on qubits to modify the qubit states.
For example, \textit{Pauli} gate is for a single qubit, and \textit{CNOT} (Control-NOT) gate is a commonly-used quantum operation applied on two qubits simultaneously.
Like classical logic gates, where the conventional digital circuits are the basic building blocks,
quantum gates are also the building blocks of quantum circuits.



\subsection{Quantum Computing Workflow}

The general \textit{quantum workflow} includes quantum circuit implementation, quantum computer selection, and qubit mapping \& execution~\cite{weder2021automated}.

\textbf{Quantum Circuit Implementation.} Quantum circuit implementation is the first step of the quantum workflow where users implement a quantum circuit to realize a quantum algorithm. A logical quantum circuit is ready to be compiled and executed by connecting multiple qubits with quantum gates. There are many ways to implement a quantum circuit. For example, the IBM Quantum platform supports circuit implementation by an online graphical interface, and the implementation by programming with \textit{Qiskit}~\footnote{Qiskit: \url{https://qiskit.org/}} is also supported. 

\textbf{Quantum Computer Selection.} After implementing a usable logical quantum circuit, users are required to select a preferred quantum computer for further execution. This step is totally up to users to decide which quantum computer is appropriate. A constraint for quantum computer selection is the required number of qubits for a given circuit. For example, the quantum computer \textit{ibmq\_armonk} in IBM Q platform with only one qubit is not appropriate to host a circuit with five qubits required. A quantum computer with a better performance of qubits and quantum gates would be more reliable to execute a quantum circuit~\cite{ash2019qure}. In addition, the job queuing number determines the waiting time for a circuit execution. Thus, quantum computer selection is a complex task.

\begin{figure}[t]
\centering
\includegraphics[width=0.9\columnwidth]{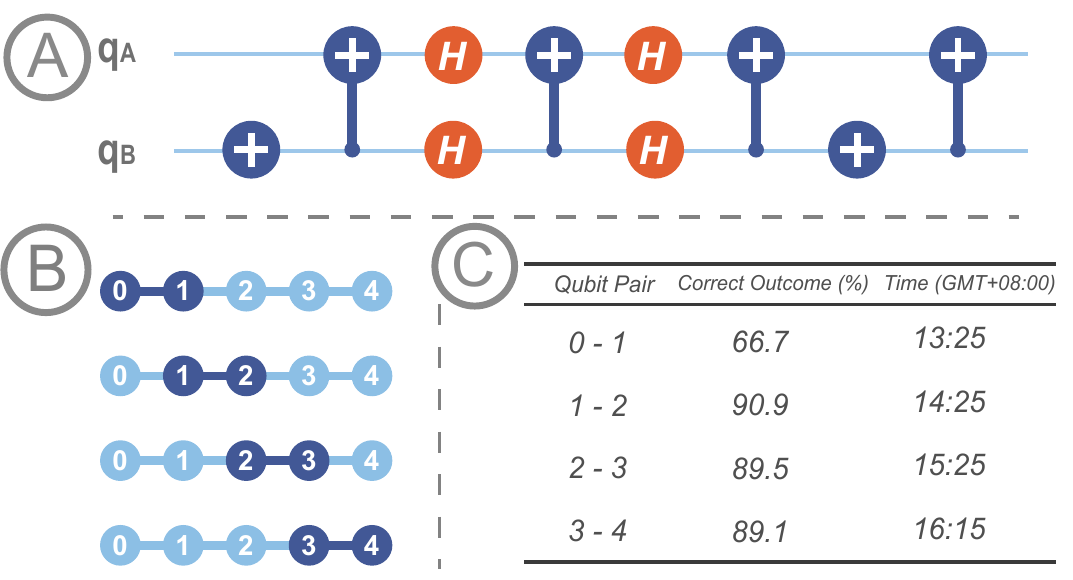}
\caption{An example of qubit mapping on quantum computer \textit{ibmq\_bogota}. (A) the test logical circuit.  \modify{The symbol ``+'' indicates the CNOT gate, and ``H'' indicates the Hadamard gate which creates the superposition states.} (B) Five options of qubit mapping on physical qubits of \textit{ibmq\_bogota}. No SWAP gate is needed due to the direct connection for all options. (C) Correct output probabilities for all mapping options. 
The experiment has been conducted on January-8-2022.}
\label{fig:1}
\end{figure}

\textbf{Qubit Mapping \& Execution.} As mentioned in Section 1,
the logical circuit
will be mapped to a ``physical quantum circuit''~\cite{venturelli2019quantum} to fit the connectivity of the selected computer.
The qubit mapping procedure depends on the mapping algorithms and cannot be modified by users. The mapped physical circuit can be different even for the same logical circuit~\cite{ venturelli2019quantum}.
Fig. \ref{fig:1}\component{B} shows four different mapped physical circuit options for the same logical circuit (Fig. \ref{fig:1}\component{A}). Since the mapping is a blind mapping without any noise awareness, the probability of the correct results would be different due to the performance variation of qubits and quantum gates~\cite{ash2019qure} (Fig. \ref{fig:1}\component{C}).

\subsection{Fidelity \& Noise in Quantum Computing}

\toolName\ uses the \textit{fidelity} of the execution results to assess the \textit{qubit error} and \textit{gate error} in quantum computers and compiled circuits.

\textbf{Fidelity.} Fidelity is a metric for measuring the difference between ideal execution results (noise-free) and actual results. For today's NISQ computer, the execution of a quantum algorithm is error-prone, leading to biased results of measurements compared with a noise-free result. The higher the fidelity is, the more accurate and valid an execution is. 

\textbf{Qubit Error.} A qubit can retain the state for only a limited period of time, and this duration is called \textit{decoherence time}. There are two metrics to measure the decoherence error, \textit{i.e.,} \textit{relaxation time T1} and \textit{dephasing time T2}. 
T1 affects the state $\ket{1}$ (i.e., $\ket{1} \to \ket{0}$), leaving state $\ket{0}$ invariant. However, it is also possible that qubit may interact with the environment and encounter a phase error, and the time constant associated with this decay is called dephasing time T2. 
Besides the decoherence error, errors can also exist in the readout of the qubit state, which is called the \textit{readout error}~\cite{bjfa, tannu2019not}.

\textbf{Gate Error.} Gates are operations controlling multiple qubits. They can also affect qubit states due to the noise.
Gate error rate is defined as the probability of introducing an error while performing operations~\cite{knill2008randomized}. Different types of gates may have different impacts on fidelity. For example, a recent study reveals the error rate of a 2-qubit gate (\textit{e.g.}, CNOT gate) can cause the maximum loss of fidelity~\cite{ash2019qure}. 


\section{Informing the design}

We conducted a pilot study to identify the general workflow and derive design requirements for noise awareness in quantum computing.

\subsection{Pilot Study}
\label{sec-pilot-study}

We designed a pilot study following the methodology introduced by Sedlmair \textit{et al.}~\cite{sedlmair2012design}.
Specifically, we invited five quantum computing experts (\textbf{P1-5})
to participate in
the pilot study. 
\modify{\textbf{P1} is a research scientist from Pacific Northwest National Laboratory,
\textbf{P2-4} are professors or post-doc researchers from three different universities, and \textbf{P5} is a Ph.D. student majoring in quantum computing.
Among them,
\textbf{P1} and \textbf{P2} are doing research on quantum chemistry.
\textbf{P3} is working on quantum machine learning, and is one of the co-authors of this work.
\textbf{P4} and \textbf{P5} study the temporal patterns of quantum hardware's quality.
All the domain experts have an average of 4.9 years of research and development experience in quantum computing.}

We began the pilot study by performing one-on-one, semi-structured, hour-long 
interviews with all participants. We categorized all participants into two groups. First, Group 1 (\textbf{P1-3}) were asked several questions about their understandings and solutions to noise awareness in quantum computing. 
Based on their feedback, we formulated the initial design requirements and proposed a low-fidelity demo system. 
Next, we presented the demo system to Group 2 (\textbf{P4-5}), who can provide in-depth suggestions for the noise analysis based on their calibration data analysis experiences of quantum computers.
We discussed the demo system with Group 2 iteratively. They were asked several questions about their suggestions and concerns; we tuned the demo system accordingly. We observed and took notes during the meetings to ensure that our early system meets the basic requirements for the quantum computing scenario. Second, we presented the early system and held weekly meetings with Group 2 over the next two months.
They were asked to explore the prototype system freely and complete the analysis tasks for noise-aware quantum circuit execution. We recorded each meeting and then held open discussions separately. The suggestions collected in this round were used to polish our system further.






\subsection{Design Requirements}

According to the feedback from all the experts, there are two major steps for executing a quantum algorithm on a quantum computer: \textbf{quantum computer selection} and \textbf{compiled circuit selection}.
Even on the same quantum computing platform, different quantum computers can have significantly different hardware properties such as gate-count, qubit-count, performance data, and qubit connectivity.
Also, selecting a better compiled circuit can improve the fidelity and reliability of execution due to the noise variation of the compiled circuits.
We summarized six design requirements for the evaluation and mitigation of noise.

For the quantum computer selection, users need to select the optimal quantum computer to host the quantum algorithm, which requires users to be aware of quantum noise from the following aspects:




\begin{itemize}
  \setlength\itemsep{0pt}
  
    \item[\textbf{R1}] \textbf{Facilitate the temporal analysis of qubit and gate noise.}
    All participants (\textbf{P1-5}) confirmed that it is crucial to enable the temporal exploration of the qubit and quantum gate noise. Specifically, the temporal analysis of quantum computers' status will be more accurate than the reflection under a single timestamp. \textbf{P1} also commented that it will be intuitive to reveal the periodical patterns of qubits' performance~\cite{bjfa} via a visualization approach.

    \item[\textbf{R2}] \textbf{Make users aware of the latest quantum computing noise.} 
    \textbf{P1} and \textbf{P5} confirmed that besides tracing the historical noise of quantum computers, it is also important to be informed of the latest noise status, since the time-varying quantum hardware properties can change quickly when the calibration launches. \textbf{P4} also reported that the real-time reflection of the queuing job number will also be helpful for quantum computer selection.
    
\end{itemize}

For the compiled circuit selection, the following requirements are crucial for informing the quantum noise users of different noise:

\begin{itemize}
  \setlength\itemsep{0pt}

    \item[\textbf{R3}] \textbf{Provide an overview of all compiled circuits.} According to the suggestions from four participants (\textbf{P2-5}), more compilation results can increase the likelihood of selecting an optimal compiled circuit. Thus, it will be helpful to allow users to select multiple compiled circuits of their interest for an in-depth comparison.

    \item[\textbf{R4}] \textbf{Enable a detailed comparison of the usages of implemented qubits and gates.} All participants (\textbf{P1-5}) emphasized the need for a drilling-down comparison of selected compiled circuits.
    In addition, \textbf{P4} also pointed out that it is necessary to provide the reference value of the performance for a better noise comparison.

    \item[\textbf{R5}] \textbf{Support a real-time compilation and fidelity validation of quantum algorithms.} All participants (\textbf{P1-5}) confirmed that the system should support real-time compilation of quantum algorithms on real quantum computers to generate the latest compiled circuits.
    Also, the preferred compiled circuit needs to be further executed on the selected quantum computer, which can help users validate the performance of the selected quantum computer and compiled circuits for running the selected algorithm.
    
\end{itemize}

For the overall visual designs, experts also requested flexible interactions and intuitive visualizations:
\begin{itemize}
  \setlength\itemsep{0pt}
    
    \item[\textbf{R6}] \textbf{Provide flexible user interactions and intuitive visual designs.}
    Three participants (\textbf{P1}, \textbf{P3-4}) commented that the system needs to support on-demand analysis for quantum computing noise assessment. For example, some casual users will focus on the quantum gate error rate to handle simple quantum circuits,
    while other users may be interested in the qubit decoherence time when executing a complex quantum circuit like a quantum machine learning model. Flexible user interactions can support such requirements. 
    Also, concise design is preferred, as quantum computing users often do not have a background in data visualization.
\end{itemize}

\subsection{Dataset}
\label{subsec:dataset}


Guided by the above design requirements, we collected the following data from the cloud quantum platform (\textit{i.e.,} IBM Quantum), which indicates quantum noise and will be visualized in \toolName{}:

\begin{itemize}
\setitemize{noitemsep,topsep=0pt,parsep=0pt,partopsep=0pt}
    \item \textbf{Calibration data of quantum computers.} We collected the latest relaxation time T1, dephasing time T2, qubit readout error, and gate error rate. We also collected the queuing number of the quantum computer from the remote quantum cloud computing platform.
    
    \item \textbf{Properties of compiled physical circuits.} Such properties need to be extracted from real-time compilation, which includes the usage of each qubit and quantum gate, and the execution results of a quantum algorithm from the cloud quantum computing platform. Following the prior studies~\cite{resch2020day, gokhale2020optimized, wille2019mapping}, we further calculate the fidelity of each compiled physical circuit by using the Hellinger distance~\cite{luo2004informational}. 
\end{itemize}

\definecolor{module_ibmq}{RGB}{167, 87, 173}
\definecolor{module_storage}{RGB}{127, 156, 214}
\definecolor{module_processing}{RGB}{234, 156, 78}
\definecolor{module_interface}{RGB}{80, 170, 119}

\begin{figure}[t]
\centering 
\includegraphics[width=\columnwidth]{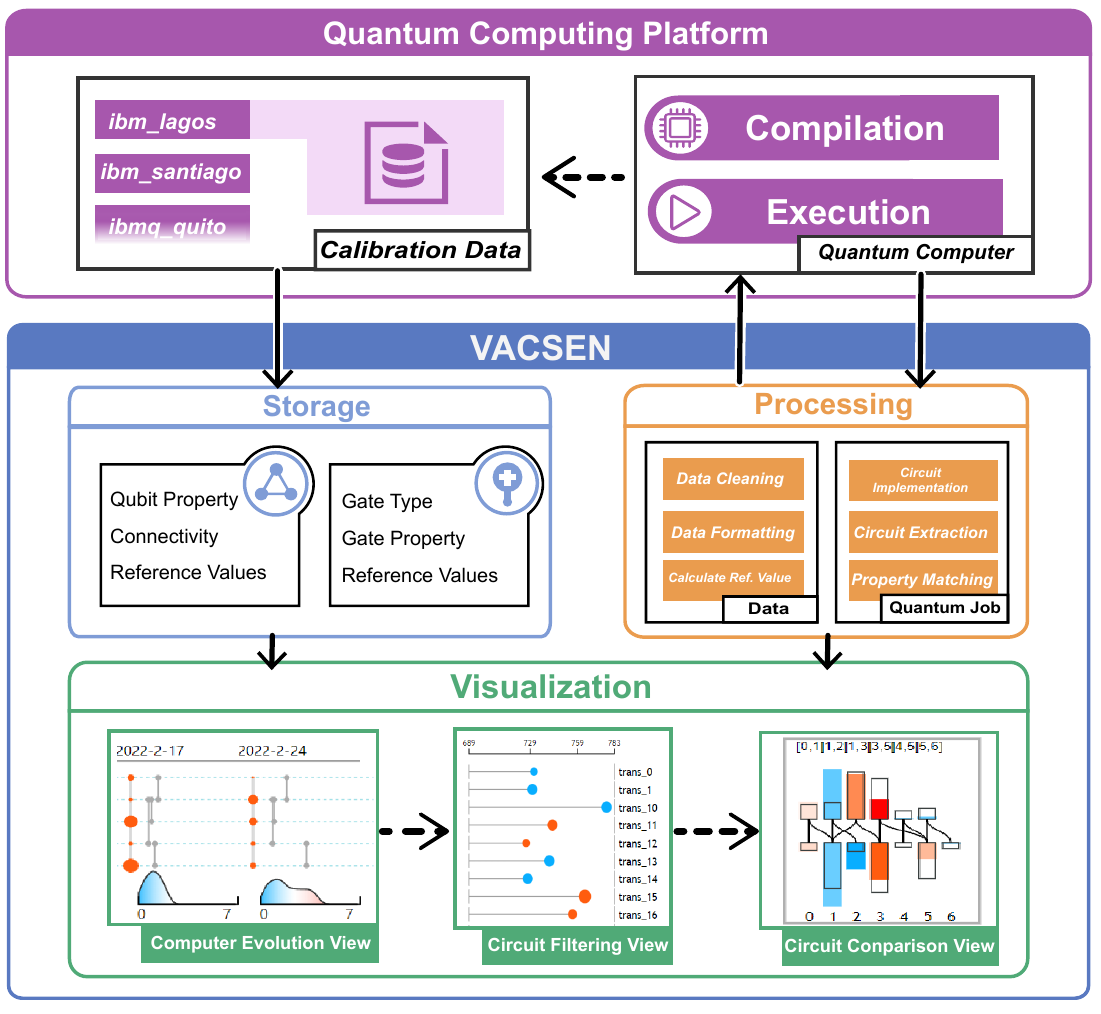}
\caption{The system architecture of \toolName{} contains three modules (a \textcolor{module_storage}{\textbf{storage}} module, a \textcolor{module_processing}{\textbf{processing}} module, and an \textcolor{module_interface}{\textbf{visualization}} module) and the remote \textcolor{module_ibmq}{\textbf{quantum computing platform}}.}
\label{fig:2}
\end{figure}

\section{\toolName}
\label{sec:tech}

We propose \toolName{}, an interactive visualization approach to inform users of the noise in quantum computing when trying to execute their quantum algorithms on quantum computers.
\toolName\ can be accessed via the URL: \textcolor{blue}{\url{https://vacsen.github.io/}}.


Fig. \ref{fig:2} shows the architecture of \toolName.
\toolName{} is designed to work with quantum computing cloud platforms (\textit{e.g.}, IBM Quantum),
and inform users of the quantum noise in
the quantum circuit compilation and execution.
\toolName{} consists of three modules: storage module, processing module and visualization module.
The storage module stores all the available 
latest calibration data, including qubit properties and gate properties.
The processing module handles the quantum algorithm's implementation and the compiled circuit's extraction from the cloud quantum computing platform.
The visualization module reveals the noise of quantum computers and compiled circuits.

\modify{After selecting the optimal compiled circuit, \toolName\ also supports a post-execution analysis for a quantum algorithm.
Fidelity Comparison View (Fig. \ref{fig:teaser}\component{E}) supports the validation of the compiled circuits' fidelity.
The vertical coordinate of the dot indicates the fidelity value.
The dot highlighted in black in Fidelity Comparison View
denotes the fidelity value of the compiled circuit selected by the user.
Also, users can further check their detailed state distributions by clicking the dot.
The corresponding execution result will be visualized on Probability Distribution View (Fig. \ref{fig:teaser}\component{F}), where the x-axis indicates the state distribution and the y-axis represents the corresponding shot numbers.
}

\subsection{Computer Evolution View}
\label{subsec:temporal_view}


\begin{figure}[t]
\centering 
\includegraphics[width=\columnwidth]{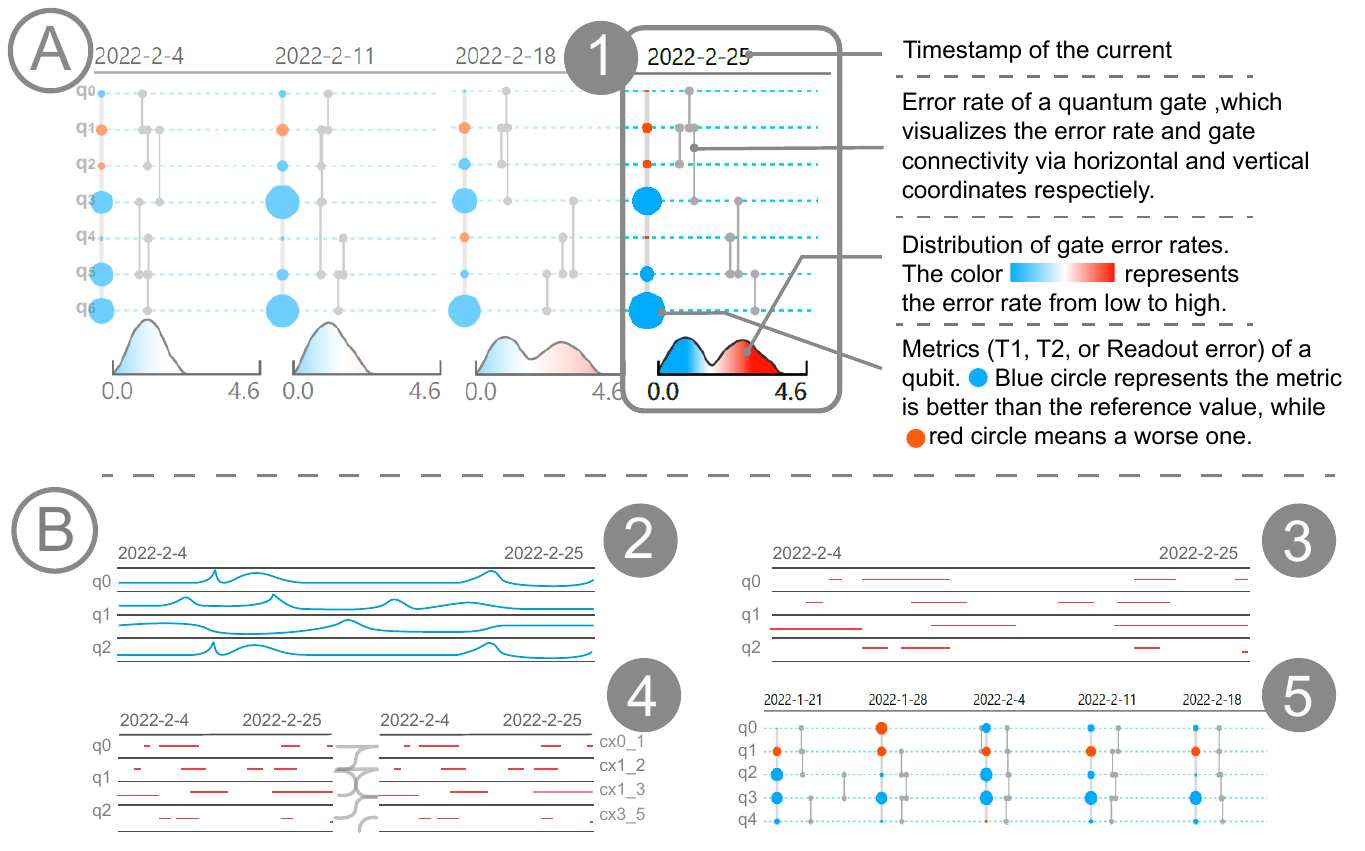}
\caption{Computer Evolution View. (A) The time-varying quality of all usable quantum computers are visualized using the uniform timeslicing approach. (B) Design alternatives of the Computer Evolution View.
\modifyRed{
The alternative designs (2) and (3) cannot visualize the connections among qubits. While the other alternative designs either suffer from visual clutters (4) in visualizing qubit connections or cannot clearly indicate gate errors (5).
}
}
\label{fig:3}
\end{figure}

We propose a circuit-like design to visualize the quantum computer noise in each timestamp. Following the prior studies~\cite{bhattacharjee2019muqut, bjfa, ash2019qure}, we encode qubit T1 time, qubit T2 time, qubit readout error rate, and quantum gate error rate to portray the noise in quantum computers. 
We utilize a sequence of circuit-like designs to visualize the time-varying performance of a quantum computer.
Also, as shown in Fig. \ref{fig:teaser}\subcomponent{A\textsubscript{6}}, the latest queuing number of the quantum computer is encoded by the horizontal bar chart.


\textbf{Circuit-like Design. }
As shown in Fig. \ref{fig:3}\component{A}, 
the colored circles represent the value of one of the qubit noise attributes (\textit{i.e.,} T1 time, T2 time, and readout error), which can be switched by the user. 
According the feedback of domain experts, the relative noise across different qubits and gates is even more effective than their absolute noise values for quantum computers and compiled circuit selection.
Thus, 
we first calculate the average of the noise attributes among all qubits as the reference value and further compute the difference between a noise attribute and the reference value.
Specifically,
the circle radius represents the absolute value of the difference between the readout error and the reference value.
\modify{A circle will be colored in blue \blueSquare \ \ if the readout error is lower than the reference value. Otherwise, it will be colored in red \redSquare \ \ . }
The larger the radius of the blue circle, the better the performance of the qubit; conversely, the larger the radius of the red circle, the worse the performance of the qubit; if the radius of the circle is close to 0, it means that the performance of the qubit is close to the average performance.
Meanwhile, the quantum gate topology is encoded by the line segments in grey. 
The horizontal coordinate of each line segment indicates its gate error rate, while the vertical position of two endpoints denotes the two qubits that the quantum gate operates on.  
We adjust the opacity of line segments to mitigate the possible visual clutter issues due to overlapped line segments.

\definecolor{subcomponent}{RGB}{137, 137, 137}
\newcommand*\Subcomponent[1]{\tikz[baseline=(char.base)]{
            \node[shape=circle,fill=subcomponent,text=white, inner sep= 1.3pt] (char) {\textsf{\small #1}}}}

During the iterative weekly meetings with domain experts, they mentioned that it is difficult to understand the overall noise level of quantum gates. 
To this end, we visually summarized the overall distribution for quantum gates' noise using the Kernel Density Estimation (KDE) method. Given all quantum gates' error rate distribution list $(x_1, x_2, ..., x_n)$, the KDE can be calculated as follows:

\begin{equation}
\label{equation:2}
\hat{f}_h(x) = \frac{1}{nh}\sum_{i=1}^{n} K(\frac{x-x_{i}}{h}),
\end{equation}

where $K$ is the kernel - a non-negative function - and $h$ is a smoothing bandwidth. We utilize the Gaussian function~\cite{grinstead1997introduction} as the kernel of the KDE distribution.
As shown in Fig. \ref{fig:3}\Subcomponent{1}, we visualize the KDE distribution at the bottom of each circuit-like design. 
Furthermore, we fill the density area chart with a gradient color (\textit{i.e.,} from blue to red)
to intuitively show the quantum gate noise distribution.

\textbf{Design Alternatives.} 
Before finalizing the current visual design,
we also considered four design alternatives for the temporal analysis of the quantum computer noise (Fig. \ref{fig:3}\component{B}).
Fig. \ref{fig:3}\Subcomponent{2} is the initial design that is a general design for temporal analysis. However, it is not appropriate to visualize a number of qubits' noise trends simultaneously in a narrow space, making it difficult to observe the temporal patterns.
To address the problem, we abstracted the anomalous periods and highlighted them with red line segments (Fig. \ref{fig:3}\Subcomponent{3}). 
During our weekly meeting with domain experts,
all the experts (\textbf{P1-5}) commented that it would be better to embed the quantum gates' noise and their connectivity into the design. The design in Fig. \ref{fig:3}\Subcomponent{4} addressed this problem, but the curve lines denoting the gate connectivity will introduce a severe visual clutter when the qubit number is over five.
Fig. \ref{fig:3}\Subcomponent{5} addressed the challenge by using a timeslice approach. 
However, as mentioned above, the quantum gate noise distribution is not intuitive for this design.

\subsection{Circuit Filtering View}
\label{subsec:equation3}

\begin{figure}[tb]
\centering 
\includegraphics[width=\columnwidth]{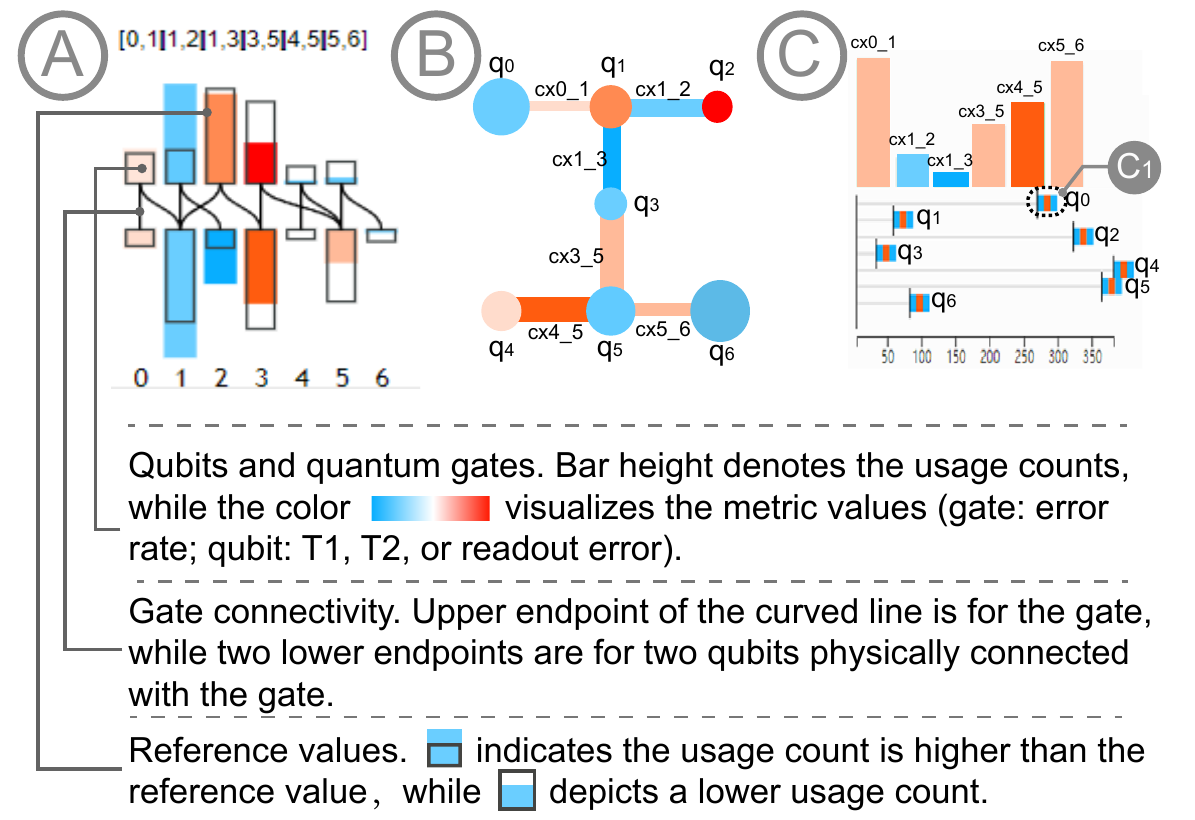}
\caption{Circuit Comparison View. (A) The Circuit Comparison View supports the in-depth comparison of multiple compiled circuits. (B)(C) Design alternatives of Circuit Comparison View.
\modifyRed{
Both alternative designs cannot support an effective comparison of the usage times of qubits and quantum gates.
}
}
\label{fig:4}
\end{figure}

The Circuit Filtering View aims to help users filter several compiled circuits of their interest.
To portray the overall performance of qubits and gates, we define a metric \textit{overall performance scores}, including \textit{qubit scores} and \textit{gate scores}, for each compiled circuit.
We then calculate the average of all scores as the reference value.
As shown in Fig. \ref{fig:teaser}\component{B}, each row denotes a compiled circuit.
The horizontal coordinate of the circle depicts the \textit{depth} of the circuit, which indicates the number of instructions for a compiled circuit. 
The circle will be colored blue if the overall performance scores are higher than the reference value. Otherwise, it would be rendered in red.
The user can configure in the control panel and indicate either gate scores or qubit scores to be visualized by the circles.
The circle radius indicates the absolute value of the difference between the overall performance scores and the reference value.
The larger the radius of the blue circle, the less noisy the circuit; conversely, the larger the radius of the red circle, the noisier the circuit; if the circle radius is closer to 0, it means that the circuit noise is closer to the average level.

\modify{For a compiled circuit, a physical quantum gate or qubit is often be used for multiple times. 
After discussing with the domain experts, we followed the idea of Linearly Weighted Moving Average and proposed the overall score ($S$), which represents the weighted error rate across different qubits or gates to indicate the overall error rate of a compiled circuit.
We choose the reciprocal of the overall error rate to represent its overall performance.
Thus, given a compiled circuit with $N$ quantum gates (or qubits), we define the overall score $S$ as follows:}

\begin{equation}
\label{equation:4}
S = \modifyRed{(}\frac{\sum_{i=1}^{N}{C_{i} \cdot E_{i}}}{\sum_{i=1}^{N}{C_{i}}}\modifyRed{)}^{\modifyRed{-1}},
\end{equation}

\modify{where $E_{i}$ denotes the error rate of a quantum gate (or qubit), and $C_{i}$ is the usage times of a quantum gate (or qubit). }

\subsection{Circuit Comparison View}

Circuit Comparison View supports the in-depth comparison of multiple compiled circuits.
We propose a novel coupled bar chart to enable users
to drill down to the previously selected compiled circuits. 
The design summarizes the instructions of the compiled circuits. It portrays the noise based on the performance metrics (\textit{i.e.,} qubit T1 time, qubit T2 time, qubit readout error, and quantum gate error rate) and the usage times of the hosted qubits and quantum gates.

\textbf{Coupled Bar Charts.} 
As shown in Fig. \ref{fig:4}\component{A}, 
each coupled bar chart indicates the noise for a compiled circuit.
The upper group of bars denotes the usage times of quantum gates of a compiled circuit, while the bottom group of bars represents the usage times of the qubits. 
The color of the upper bars represents the error rate of the gates, while the color of the lower bars denotes one of the three noise attributes for qubits (\textit{i.e.,} T1, T2, and readout error).
The darker the blue of the bar, the less the noise of the qubit or gate; the darker the red of the bar, the more noise of the qubit or gate.
The height of the black rectangle indicates the average usage times of the gate or qubit among all compiled circuits.
Meanwhile, the curved lines between the upper and lower groups of bars visualize the connectivity of quantum gates. Specifically, the upper endpoint of the curved line represents a quantum gate, while the corresponding two lower endpoints denote the two qubits operated on the quantum gate. 

\textbf{Design Alternatives.} 
We considered other two design alternatives for the detailed comparison of compiled circuits regarding the noise. 
Fig. \ref{fig:4}\component{B} shows the compiled circuit's noise by the original topology of the selected quantum computers. The usage times of qubits and quantum gates are encoded by the circle radius and line segment width, respectively. The noise level is indicated by the same color encoding as the coupled bar chart. However, it is difficult to compare the line segment width between multiple compiled circuits. 
The second design alternative is shown in Fig. \ref{fig:4}\component{C}, which encodes the gate usage times by the bar height and qubit usage times by the horizontal position of the glyph (Fig. \ref{fig:4}\subcomponent{C\textsubscript{1}}). However, we found that the comparison between multiple compiled circuits from horizontal and vertical directions simultaneously is confusing for the users.
To address all these challenges, we further propose the coupled bar chart (Fig. \ref{fig:4}\component{A}).

\subsection{User Interactions}

\toolName{} enables rich interactions to help users smoothly explore and analyze the noise of different quantum computers and compiled circuits.

\textbf{Hierarchical Noise Analysis.}
The Computer Evolution View (Fig. \ref{fig:teaser}\component{A}) provides an overview of quantum computer noise.
By clicking a quantum computer, users can view the noise of all the compiled circuits in the Circuit Filtering View (Fig. \ref{fig:teaser}\component{B}), which can be further explored and compared in detail in the Circuit Comparison View (Fig. \ref{fig:teaser}\component{C}).

\textbf{Time Configurations.}
Users can adjust the time range of the noise to be shown in the Computer Evolution View by specifying the latest \textit{``time range''} value in the control panel (Fig. \ref{fig:teaser}\component{D}) and the time \textit{``interval''}~(Fig. \ref{fig:teaser}\component{D}) can also be adjusted to show quantum computer noise with different granularity.


\textbf{Quantum Algorithm Specifications.}
Users can specify the quantum algorithm to be run and the number of compilations in the control panel (Fig. \ref{fig:teaser}\component{D}).
Once the quantum algorithm is determined, users can launch the compilation of the quantum algorithm on the cloud platform by simply clicking the button \textit{``Compile''} (Fig. \ref{fig:teaser}\component{D}).


\textbf{Interactive Filtering and Sorting of Compiled Circuits.}
Users can switch between ``Gate'' and ``Qubit'' to visualize the compiled circuits according to either gate scores or qubit scores.
Also, \toolName{} allows users to sort all compiled circuits in the Circuit Filtering View by the quality scores or circuit depth via the switch ``\textit{Sort by scores}'' and ``\textit{Sort by depth}'' in the control panel (Fig. \ref{fig:teaser}\component{D}).
Furthermore, users can filter circuits by their quality scores via the ``Score Filter'' range slider.


\textbf{Remote Execution Controlling.}
After selecting the preferred compiled circuit for execution,
the user can launch the execution via the button ``\textit{Run}'' in the control panel (Fig. \ref{fig:teaser}\component{D}). The remote cloud quantum computer would host and execute the selected compiled circuit.



\section{Case Study}

In this section, we conducted two case studies on both small-scale and large-scale quantum circuits to demonstrate the effectiveness of \toolName{}.
The users involved in the case studies are two quantum computing experts (U1 and U6)
who also attended the user interviews
in Section~\ref{sec:user-interview}.
IBM Quantum platform was used to compile and execute the quantum circuits on March 5, 2022\footnote{
Only nine out of 10 quantum computers are in service on March 5, 2022 (the date of our case studies), while the unavailable quantum computer \textit{ibmq\_santiago} is under maintenance}.

\subsection{Case Study \uppercase\expandafter{\romannumeral 1} - Two-qubit Circuit}
\label{subsec:case_2}

U1 employed \toolName{} to explore the noise of different quantum computers and select an appropriate compiled circuit for the two-qubit circuit introduced by Ash-Saki \textit{et al.}~\cite{ash2019qure}.
This circuit is often used for demonstrating the different probabilities of correct results.

\begin{figure}[t]
\centering 
\includegraphics[width=\columnwidth]{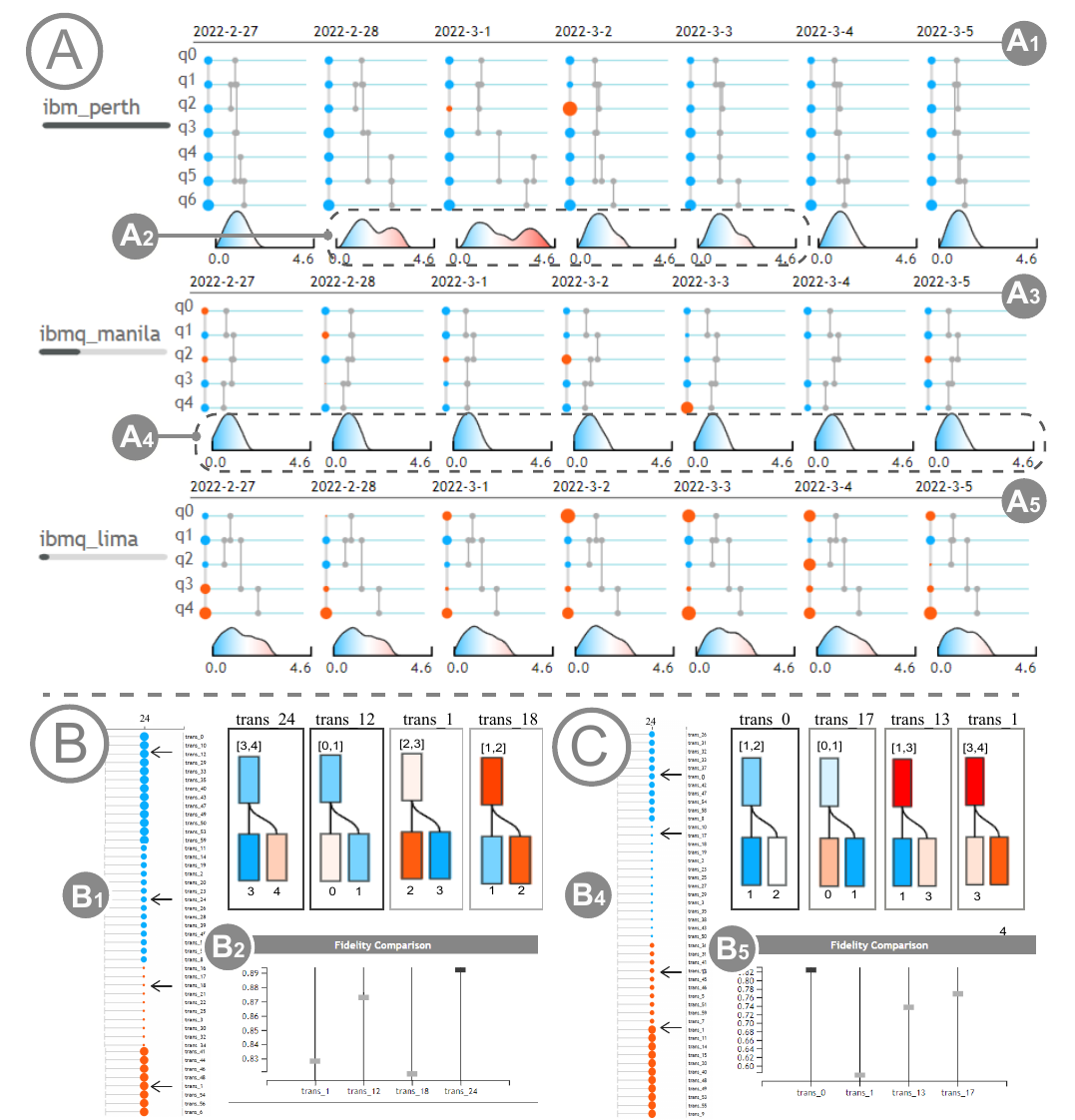}
\caption{The case for two-qubit circuit. (A) Computer Evolution View for temporal analysis of \textit{ibm\_perth}, \textit{ibmq\_manila}, \textit{ibmq\_lima}. (B) Comparison of compiled circuits for \textit{ibmq\_manila}. Comparison of compiled circuits for \textit{ibmq\_lima}.}
\label{fig:5}
\end{figure}

\textbf{Striking a trade-off between quantum computer noise and queuing time.} 
Both quantum computer noise assessment and queuing number awareness are crucial for quantum computer selection, which can make the execution more reliable and time-saving.
Given that the quantum circuit needs two qubits, eight potential quantum computers can be used, except \textit{ibmq\_armonk} with only one qubit.
U1 first compared the noises of different quantum computers in the past week (\textbf{R1}) and set the time interval as one day in the control panel.
U1 then set the \textit{``Qubit Noise''} as the readout error (\textbf{R6}).
By simply glancing at all the available quantum computers,
U1 immediately noticed that \textit{ibm\_perth} has the least qubit noise (\textit{i.e.,} the best qubit performance) from 2022-2-27 to 2022-3-5, as indicated by the consistently large blue circles (Fig. \ref{fig:5}\subcomponent{A\textsubscript{1}}).
The gate error rate of \textit{ibm\_perth} was rather low on the current day (\textit{i.e.,} 2022-3-5), though it fluctuated from 2022-2-28 to 2022-3-3 as indicated by the density area charts (Fig. \ref{fig:5}\subcomponent{A\textsubscript{2}}). 
He then noticed the quantum computer \textit{ibmq\_manila}, whose gate error rates have been stable in the past week, staying below 1\% (Fig. \ref{fig:5}\subcomponent{A\textsubscript{4}}). Meanwhile, the qubit noise of \textit{ibmq\_manila} is quite low via inspecting the blue circles.
Thus, U1 concluded that the noise of quantum computers  \textit{ibm\_perth} and \textit{ibmq\_manila} is the least among all quantum computers. 
U1 evaluated \textit{ibm\_perth} and \textit{ibmq\_manila} to the same noise level regarding the qubit and gate errors.
Furthermore,
U1 noticed that the queuing numbers for \textit{ibm\_perth} (i.e., 430 in total) were much larger than other computers, as indicated by the bar charts below the computer names.
Given that the waiting time of \textit{ibm\_perth} is a bit long, 
U1 finally selected \textit{ibmq\_manila} for the subsequent quantum circuit execution.

\textbf{Unveiling the mask of low-fidelity compiled circuits.} 
After selecting the appropriate quantum computer (\textit{i.e.,} \textit{ibmq\_manila}), U1 then moved on to the next step (\textit{i.e.,} quantum circuit compilation). 
U1 further used \toolName{} to filter the suitable compiled circuits on \textit{ibmq\_manila}.
U1 first changed the number of compilation times to 60 in the control panel.
After a few seconds of compilation, all the compiled circuits were displayed in the Circuit Filtering View.
Then, U1 sorted all the compiled circuits according to the gate scores (Fig. \ref{fig:5}\subcomponent{B\textsubscript{1}}).
U1 noticed that the 60 compiled circuits have the same depth of 24, as indicated by the circles' horizontal coordinates. 
Meanwhile, from the radii of all the circles, U1 speculated that the two-qubit logical circuit can be compiled into four different physical circuits based on the qubit mapping algorithm.
U1 randomly selected one compiled circuit from each category and
switched to the Circuit Comparison View
for a detailed comparison.
U1 immediately noticed that the four circuits are assigned on different physical quantum gates on \textit{ibmq\_manila}, as indicated by the four coupled bar charts in Fig. \ref{fig:5}\component{B}.
For example, Circuit \textit{trans\_24} with the highest gate scores assigned the CNOT gate on Gate \textit{cx3\_4} and Qubits \textit{q3} and \textit{q4}, whose gate error is minimal as indicated by the blue bar on the top.
Also, U1 found that the two bars in Circuit \textit{trans\_18} were in dark red, indicating that \textit{trans\_18} utilized Gate \textit{cx1\_2} and Qubit \textit{q2} with the highest error rate for the quantum circuit implementation.
Thus, U1 decided to select Circuit \textit{trans\_24} for the final execution.
To test the effectiveness of \toolName\ and make sure Circuit \textit{trans\_24} was the circuit with the least noise, U1 also executed the other three compiled circuits. 
Fig. \ref{fig:5}\subcomponent{B\textsubscript{2}} shows the fidelity distribution for the four compiled circuits. It was apparent that the preferred compiled circuit \textit{trans\_24} represented by the black rectangle had the highest fidelity (\textbf{89.4\%}). In contrast, the fidelity of the other three compiled circuits was \textbf{87.5\%}, \textbf{83\%} and \textbf{82.1\%}, respectively. 
U1 finally got the best execution results with the least noise with the help of \toolName.

\textbf{Selecting the best of a bad bunch.} 
U1 planned to execute the two-qubit circuit on another  quantum computer with small queuing numbers to see if it is possible to generate a high-fidelity result with little queuing time. As shown in the Computer Evolution View (Fig. \ref{fig:5}\subcomponent{A\textsubscript{5}}), the performance of \textit{ibmq\_lima}'s qubits was below average as indicated by the red circles, and several gates (Gate \textit{cx1\_3} and Gate \textit{cx3\_4}) also performed badly in the past week according to their line segments' horizontal coordinates. However, due to the small queuing number of the quantum computer \textit{ibmq\_lima}, as indicated by the bar chart below the computer name, U1 planned to execute on the noisy quantum computer \textit{ibmq\_lima} to get a reliable result in a short waiting time. After the circuit compilation and sorting of all the compiled circuits in the control panel, U1 found four types of compiled circuits from Circuit Filtering View (Fig. \ref{fig:5}\subcomponent{B\textsubscript{4}}). 
U1 found that Circuit \textit{trans\_0} had the least noise due to the color of bars for the implemented Gate \textit{cx1\_2} and Qubits \textit{q1} and \textit{q2}.
Thus, U1 selected the Circuit \textit{trans\_0} for the final circuit execution.
To convince himself, U1 then executes other three compiled circuits.
The fidelity of the execution is shown in Fig. \ref{fig:5}\subcomponent{B\textsubscript{5}}. Circuit \textit{trans\_0} showed a much better fidelity (\textbf{83.1\%}) than the other circuits (\textbf{77.5\%}, \textbf{74.4\%} and \textbf{58.5\%}).


\subsection{Case Study \uppercase\expandafter{\romannumeral 2} - Shor's Algorithm}

U6 used \toolName{} to assess the quantum noise when trying to execute a large-scale quantum circuit, \textit{i.e.,} \textit{Shor's algorithm}. Shor's algorithm~\cite{shor1999polynomial} is a famous and widely-used quantum algorithm for integer factorization. Meanwhile, it is usually used for the evaluation of noise optimization algorithms~\cite{salm2020nisq, de2017quantum}. 
The experiment was conducted on the IBM Quantum platform on March 5, 2022.


\textbf{Balancing different noises of a quantum computer.} 
Since the Shor's algorithm requires seven qubits, the potential quantum computers are those with no less than seven qubits, \textit{i.e.,} \textit{ibm\_lagos}, \textit{ibm\_perth} and \textit{ibmq\_jakarta}. 
U6 changed the \textit{time range} and \textit{interval} to 30 and 7 respectively in the control panel, to explore their status in the past month. 
U6 first set the qubit noise to ``\textit{readout error}'' in the control panel.
U6 quickly found that \textit{ibmq\_jakarta} (Fig. \ref{fig:teaser}\subcomponent{A\textsubscript{3}}) has the most stable gate noise from the density area charts (Fig. \ref{fig:teaser}\subcomponent{A\textsubscript{7}}), while \textit{ibm\_perth}'s gate error rate was much higher than the other two computers due to the three red areas from 2022-1-19 to 2022-2-12. 
Particularly, the error rate of Gate \textit{cx4\_5} (3.64\%) on 2022-2-12 was significantly higher than that of the other gates (Fig. \ref{fig:teaser}\subcomponent{A\textsubscript{5}}). 
Since the decoherence time matters for large-scale quantum algorithms~\cite{ash2019qure}.
Thus, U6 changed the qubit noise to ``\textit{T1}'' to assess quantum computers regarding the decoherence time.
As shown in Fig. \ref{fig:teaser}\subcomponent{A\textsubscript{4}}, 
U6 noticed that Computer \textit{ibm\_perth}, whose gates are most noisy among all the three computers as illustrated above, has the best performance for T1 time as indicated by the large blue circles. 
Meanwhile, Computer \textit{ibm\_lagos}'s T1 time is the shortest among the three computers, though \textit{ibm\_lagos} has the best performance in terms of gate error rate as indicated by the consistently blue circles in Fig. \ref{fig:teaser}\subcomponent{A\textsubscript{1}}.
Thus, U6 concluded that the three quantum computers have diverse noise patterns.
Taking into account all the factors, \textit{i.e.,} the qubit noise, gate noise, and queuing number (Fig. \ref{fig:teaser}\subcomponent{A\textsubscript{6}}), U6 decided to select Computer \textit{ibmq\_jakarta} for the final execution.

\textbf{Breaking the tie: ``similar'' compiled circuits can be different.}
After selecting \textit{ibmq\_jakarta} as the preferred quantum computer, U6 launched the compilation to map Shor's algorithm on this computer. As shown in Fig. \ref{fig:teaser}\component{B}, the depths of all the compiled circuits were from 706 to 817, which were larger compared with the previous two-qubit quantum circuit. 
Meanwhile, U6 found that every circle has a different radius, indicating that a new compiled circuit was generated at every time of the compilation. 
Since minimizing the depth of the circuit can minimize the decoherence noise for a large-scale circuit~\cite{li2019tackling, zulehner2018efficient},
U6 regarded circuit depth as the major factor for the compiled circuit selection and sorted all the 60 compiled circuits by depth in an ascending order via the switch ``\textit{Sort by depth}'' in the control panel (Fig. \ref{fig:teaser}\component{D}). 
U6 then clicked the top five circuits with a minor depth (Fig. \ref{fig:teaser}\subcomponent{B\textsubscript{1}}) for the comparison. 
Then, the five corresponding coupled bar charts were shown in Fig. \ref{fig:teaser}\component{C}.
For the gate error rates, U6 noticed that Gate \textit{cx1\_2}, which is highlighted by the purple dotted box in all the five circuits, has the highest error rate as indicated by the dark red bar. 
U6 then found that \textit{trans\_19}, \textit{trans\_40}, and \textit{trans\_5} used Gate \textit{cx1\_2} much less frequently than the other two compiled circuits (\textit{i.e.,} \textit{trans\_6} and \textit{trans\_56}) as indicated by the small height of the three bars.
Meanwhile, the usage times of the  high-performance Gate \textit{cx3\_5} in the above three circuits were much larger than those in the other two circuits, as indicated by the light blue bar's height.
U6 commented,  ``\textit{Because the extra usage numbers of the high-nois gate \textit{cx1\_2} were transferred to the low-noise gate \textit{cx3\_5}, making it more reliable.}''
For the qubit noise comparison, U6 noticed that Qubit \textit{q6}, highlighted by the purple box, has the highest noise among all qubits as indicated by the dark red bars. 
Also, it was apparent that Circuit \textit{trans\_5} used Qubit \textit{q6} most often as indicated by the bar heights, which was far above the reference value as highlighted by the black box. 
Thus, U6 decided to execute Circuit \textit{trans\_19} due to its smaller depth than \textit{trans\_40}, though they have a similar noise level.
After waiting for a while, the system displayed the fidelity distribution (Fig. \ref{fig:teaser}\component{E}), showing that the fidelity of Circuit \textit{trans\_19} (\textbf{73.6\%}) was the highest among the five circuits, as U6 expected. 
Meanwhile, U6 found that the result fidelities of \textit{trans\_19} and \textit{trans\_40} were at the first tier of less-noise circuits (Fig. \ref{fig:3}\subcomponent{C\textsubscript{1}}) due to the similar visual evidence of the corresponding coupled bar charts, while the average fidelity for the other three circuits was \textbf{58\%}.


\section{User Interviews}
\label{sec:user-interview}

To further evaluate the effectiveness of \toolName, we conducted in-depth user interviews with the target users working on quantum computing. 

\subsection{Study Design}

\begin{table}[tb]
\centering
\begin{tabular}{c|p{0.8\columnwidth}}
\hline
\modify{T1} & Find the best-quality quantum computer regarding qubit's relaxation time \textit{T1}.                         \\ \cline{2-2} 
\modify{T2} & Find the best-quality quantum computer regarding qubits’ dephasing time \textit{T2}.                         \\ \cline{2-2} 
\modify{T3} & Find the best-quality quantum computer regarding qubits’ readout error.                   \\ \cline{2-2} 
\modify{T4} & Find the best-quality quantum computer regarding gates’ error rate.                       \\ \cline{2-2} 
\modify{T5} & According to the tasks above, find the most suitable computer for the further execution. \\ \hline
\modify{T6} & Find the circuits of interest regarding the quality of building blocks.                     \\ \cline{2-2} 
\modify{T7} & Find the circuits of interest regarding the circuit depth.                                \\ \cline{2-2} 
\modify{T8} & Compare and highlight the compiled circuits with good gate-quality for the final execution.         \\ \cline{2-2} 
\modify{T9} & Compare and highlight the compiled circuits with good qubit-quality for the final execution.        \\ \hline
\end{tabular}
\caption{All tasks are grouped by the analysis workflow, \textit{i.e.,} quantum computer selection and compiled circuit selection.} 
\label{table:1}
\end{table}

\textbf{Participants and Apparatus.} We invited 12 participants (2 females) from six different educational institutions and a national research laboratory to join our in-depth user interviews. The participants are different from the experts participated in the pilot study. Specifically, \textbf{U1-5} are postgraduate students working on quantum computing. 
\textbf{U6} is a research staff from a research laboratory for quantum machine learning. 
\textbf{U7-12} are professors with an average of over five years of research experience in the quantum computing-related research fields. 
Among them, \textbf{U8-9} 's current research direction is the qubit mapping algorithm.
To guarantee that the findings from the interviews are general for common users, none of the participants has a background in visualization or HCI. 
The user interviews were conducted through the online zoom meetings due to the COVID-19 pandemic. In addition, the participants were asked to use a monitor with a resolution of 1920 $\times$ 1080 in advance.

\textbf{Quantum Algorithms and Tasks.} We provided three common quantum algorithms for quantum circuit implementation, \textit{i.e.,} Quantum Fourier Transform (\textit{QFT}) algorithm~\cite{nielsen2002quantum} and Bernstein–Vazirani (\textit{BV}) algorithm~\cite{bernstein1997quantum} (both with the depth of up to 50) and Shor's algorithm (with the depth of over 700).
We asked participants to select the quantum algorithm of interest. 
Also, we asked participants to fulfill nine carefully-designed tasks to assess \toolName, as shown in Table \ref{table:1}.

    


\begin{table}[t]
\centering
\begin{tabular}{c|p{0.8\columnwidth}}
\hline
Q1  & The system provides enough evidence to inform the noise in quantum computer.                                       \\ \cline{2-2} 
Q2  & The system enables the exploration and comparison of the noise in different compiled circuits. \\ \cline{2-2} 
Q3  & The system can facilitate noise mitigation for the quantum circuit execution.                        \\ \hline
Q4  & The overall visual design is easy to understand.                                                                   \\ \cline{2-2} 
Q5  & The visual designs of circuit-like design are helpful for assessing the noise in quantum computers                                                                          \\ \cline{2-2} 
Q6  & The design of the coupled bar charts for compiled circuit comparison is effective.                      \\ \hline 
Q7  & The user interaction of the visualization is smooth.                    \\ \cline{2-2}
Q8  & The interaction to support the communication between the remote quantum computing platform is robust                                                                \\ \hline
Q9  & The visual analytics system is easy to use                    \\ \cline{2-2}
Q10  & The visual analytics system is easy to learn                      \\ \cline{2-2} 
Q11 & I would like to use the visual analytics system to mitigate the noise in quantum computing in the future.               \\ \cline{2-2} 
Q12 & I will recommend the visual analytics system to my colleagues working on quantum computing.         \\ \hline
\end{tabular}
\caption{The questionnaire consists of four parts: the effectiveness for quantum noise awareness (Q1-3), the visual design (Q4-6), the user interactions (Q7-8) and the usability (Q9-12). 
 } 
\label{table:2}
\end{table}

\textbf{Procedures.} 
The user interview for each participant was conducted using the online \toolName\ system. We recorded and took notes for each interview and their interaction processes. 
We first introduced the analysis workflow and the corresponding visual designs of \toolName\ to the participant. After that, we showcased an example (\textit{i.e.,} the two-qubit circuit) to better illustrate the usage of \toolName. The above tutorial lasted for about 25 minutes. 
After that, the participants were asked to accomplish the pre-defined tasks of selecting an appropriate quantum computer and an optimal compiled circuit for execution.
Upon the exploration, participants were encouraged to describe the reasons for the selection in a think-aloud manner. 
The aforementioned tasks lasted about 40 minutes. 
We also invited every participants to rate the \toolName\ system based on a 7-point Likert scale from three aspects shown in Table \ref{table:2}. 
Finally, we further conducted a post-study interview with each participants, which lasted about 25 minutes.

\subsection{Result}

We summarized all participants' detailed feedback as follows:

\textbf{The Effectiveness for Noise Awareness.} 
\modify{Most participants ($rating_{mean}=5.83, rating_{sd}=1.03$) agreed that \toolName\ can facilitate noise awareness and mitigation in quantum computing.} Four participants (\textbf{U1-3}, \textbf{U11}) highly appreciate the temporal noise analysis for quantum computers. \textbf{U11} commented, \textit{``It is beneficial that I can visually analyze the time-series
pattern of various noises.
For example, I find that the qubit readout error of \textit{ibmq\_jakarta} 's q5 and q6 is always worse than that of other qubits, which is really intuitive''}
Also, three participants (\textbf{U5-6}, \textbf{U10}) commented that the proposed workflow to support the quantum computer and compiled circuit selection is quite helpful. \textbf{U10} said, \textit{``In my daily work, I need to wait for even hours for my quantum machine learning circuit execution.
Now, \toolName\ can save me a great number of time to get a similar high-fidelity result from other quantum computers with much less queuing time.''}
In addition, all participants (\textbf{U1-12}) were very interested in the communication between \toolName\ and the remote quantum computing platform IBM Q. \textbf{U8} reported that the real-time execution makes \toolName\ even more valuable and practical compared with other theoretical approaches. 

\begin{figure}[t]
\centering 
\includegraphics[width=\columnwidth]{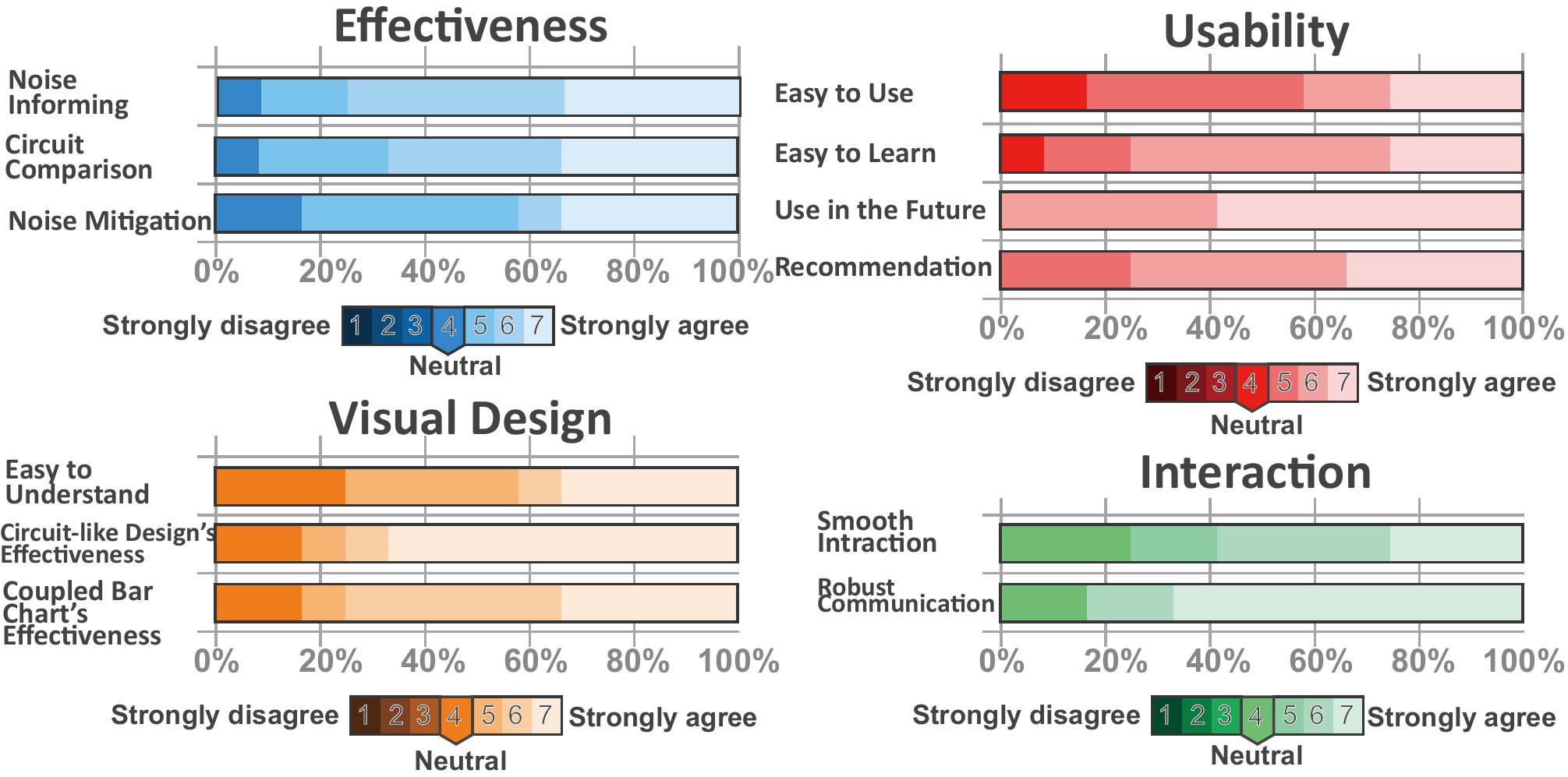}
\caption{The summary of the user feedback.}
\label{fig:6}
\end{figure}

\textbf{Usability.} 
\modify{Most participants are in favor of the usability of \toolName\ ($rating_{mean}=5.88, rating_{sd}=1.37$).} \textbf{U1-5} mentioned that \toolName\ is friendly and easy to use for quantum computing researchers. Specifically, they confirmed that \toolName\ is really easy to use. \textbf{U4} commented that the interface is really what quantum computing users want to use due to the intuitive designs.
\textbf{U6} also praised \toolName, \textit{``For quantum machine learning,
the model training device's noise state significantly impacts the model performance, and the prediction process also requires to be run on the quantum computer which has the same noise state as the one for model training.
To this end, I believe the interface is the one we are looking for to complete the noise assessments''}.
Meanwhile, \textbf{U8-9}, who is doing research on qubit mapping algorithm, commented, \textit{``I believe \toolName\ will be helpful for our current research topic of quantum network routing. We can utilize \toolName\ to host our different routing algorithm as it can reflect various noises in real-time.''}
\textbf{U5} mentioned he will utilize \toolName\ for his daily circuit execution rather than a blind execution in the past. Meanwhile, he expressed the desire to recommend \toolName\ to his colleagues in his research lab.

\textbf{Visual Design and Interactions.} 
\modify{The majority of the participants like the effective and user-friendly visual design ($rating_{mean}=6.02, rating_{sd}=0.81$) and flexible user interactions ($rating_{mean}=6.04, rating_{sd}=1.12$) of \toolName{}. }
Four participants (\textbf{U5-8}) mentioned that the circuit-like design is amazing, which encodes the topology and metrics simultaneously for each timestamp. 
Meanwhile, \textbf{U6} and \textbf{U7} commented that the density area chart within each block of Computer Evolution View is very informative. \textbf{U6} mentioned, \textit{``To my surprise, the design can visualize the distribution of all quantum gates' error rates. I can glance at the gates' density area charts and individual line segments and compare them between different quantum computers.''}
For the user interactions, \textbf{U1-12} confirmed that the system's interactions are really smooth and easy to use. 
\textbf{U9} commented,
\textit{``I found it is easy to switch between different noise attributes, which meets the needs to fit different quantum algorithm scenarios.''}

\textbf{Suggestions.} 
Despite the positive feedback, several participants also gave suggestions to improve \toolName. 
\textbf{U2} suggested that synchronous circuit execution should be more time-saving.
\textbf{U11} pointed out the selection of quantum computers can consider the correlations between various noises. 
\textbf{U10} commented that he would be happy if the \toolName\ system could support his research domain - quantum machine learning.



\section{Discussion}

In this section, we first summarize the lessons we learned during the development of \toolName. Then, we discuss the limitations of \toolName.   

\subsection{Lessons}

We learned many lessons from the system design and implementation.

\textbf{Critical importance of visualization for quantum computing.}
As shown in the above evaluations, \toolName{} received highly positive feedback from the target users. Among all the feedback, they emphasize the strong needs and critical importance of visualization approaches. 
As various noises originates from the inherent noise in current quantum computers, they really need an interactive way to investigate the reasons behind the errors of running a quantum algorithm.
\toolName{} is only the first step to address such kinds of needs in both visualization and quantum computing field, which can play an even more important role in reliable, accessible, and transparent quantum computing. 



\textbf{Intuitive visual designs matter much for quantum computing users.}
While designing the prototype of \toolName, we attempted to visualize all the noise attributes of qubits and gates simultaneously, which resulted in a sophisticated visual design.
However, during our regular meeting with the five domain experts (Section~\ref{sec-pilot-study}), they pointed out that
it is really confusing for them to explore the system because learning the 
complex visual designs
has a steep learning curve for target users. Therefore, we simplified
the visual designs, and further proposed intuitive and novel designs, such as the circuit style design and coupled bar chart
to ensure that each quantum computing user can easily use it.

\subsection{Limitations}
\label{sec:limitations}

Our evaluations have shown that \toolName\ can effectively enable noise awareness for the execution of quantum algorithms. However, the proposed approach still has limitations.



\textbf{Calibration Data Resolution.} 
\toolName\ can extract the latest calibration data to portray the performance of a quantum computer. However, according to our analysis of the calibration data, we found that the actual resolution of the data is sometimes larger than one day.
In addition, we found that the updating of the calibration data may suspend for a few days (\textit{e.g.}, \textit{ibmq\_bogota}).



\textbf{Visual Scalability.} 
\modifyRed{
According to the feedback of domain experts, general users will often
access and use only quantum computers with a relatively small number of qubits (\textit{e.g.}, up to 16) in most situations.
Our case studies and user interviews have confirmed that \toolName\ can work well for these situations.
However, \toolName\ may suffer from scalability issues when it is used to explore quantum computers with an extreme number of qubits (\textit{e.g.}, 127 or even 400).}

\textbf{Generalizability.}
\modifyRed{
\toolName\ is mainly tested using the data collected from IBM Quantum.
However, it can be easily extended to other cloud quantum computing platforms (\textit{e.g.}, Rigetti), as the workflow and performance data of quantum computers are similar across different quantum computing platforms.
}

\section{Conclusion}

We present \toolName, an interactive visual analytics approach to support noise awareness in quantum computing. 
To inform the visual design, we identified six design requirements for visually analyzing a quantum computing-specific problem. 
We proposed two novel quantum computing-specific designs, \textit{i.e.,} the \textit{circuit-like design} and \textit{coupled bar chart}, to facilitate the temporal analysis of noise in quantum computers and in-depth comparison of compiled circuits.
We conducted case studies and in-depth user interviews with 12 target users to demonstrate the effectiveness of \toolName. 
The results show that \toolName\ can effectively make the users aware of various noises and further improve the reliability of the execution result of a given quantum circuit.

In future work,
we will enhance the scalability of \toolName{} in terms of facilitating noise evaluation on quantum computers with a larger number of qubits.
\modify{Also, it will be interesting to enable automated recommendation of desirable quantum computers and compiled circuits in \toolName\ and further enhance its usability and effectiveness in helping users select the optimal quantum computers and compiled circuit. 
}


\acknowledgments{
This research was supported by the Lee Kong Chian Fellowship awarded to Yong Wang by Singapore Management University.
Qiang Guan is supported by US National Science Foundation CCF-2217021 and IBM quantum hub at NC State.
We would like to thank the participants in our user interviews and anonymous reviewers for their feedback.
}

\bibliographystyle{abbrv-doi}
\Urlmuskip=0mu plus 1mu
\bibliography{template}



\end{document}